\def\rfr#1{eq.(\ref{#1})}

\def\eqi{\begin{equation}}
\def\eqf{\end{equation}}
\def\eqia{\begin{eqnarray}}
\def\eqfa{\end{eqnarray}}
\def\rp#1#2{{#1\over#2}}
\def\asec{$''$ cy$^{-1}$}

\documentclass[onecolumn]{aa} 

\begin{document}

\title{Is it possible to measure the Lense-Thirring effect on the
orbits of the planets in the gravitational field of the Sun? }

\author{L. Iorio\inst{1}}

\offprints{L. Iorio}

\institute{Dipartimento di Fisica dell'Universit${\rm \grave{a}}$
di Bari, Via Amendola 173, 70126, Bari, Italy\\
\email{lorenzo.iorio@libero.it} }

\date{Received , 2004; accepted , 2004}

\abstract{In this paper we explore a novel approach in order to
try to measure the post-Newtonian $1/c^2$ Lense-Thirring secular
effect induced by the gravitomagnetic field of the Sun on the
planetary orbital motion.  Due to the relative smallness of the
solar angular momentum $J$ and the large values of the planetary
semimajor axes $a$, the gravitomagnetic precessions, which affect
the nodes $\Omega$ and the perihelia $\omega$ and are proportional
to $J/a^3$, are of the order of $10^{-3}$ arcseconds per century
only for, e.g., Mercury. This value lies just at the edge of the
present-day observational sensitivity in reconstructing the
planetary orbits, although the future hermean mission BepiColombo
should allow to increase it. The major problems come from the main
sources of systematic errors. They are the aliasing classical
precessions induced by the multipolar expansion of the Sun's
gravitational potential and the classical secular $N-$body
precessions which are of the same order of magnitude or much
larger than the Lense-Thirring precessions of interest. This
definitely rules out the possibility of analyzing only one orbital
element of, e.g., Mercury. In order to circumvent these problems,
we propose a suitable linear combination of the orbital residuals
of the nodes of Mercury, Venus and Mars which is, by construction,
independent of such classical secular precessions. A 1-sigma
reasonable estimate of the obtainable accuracy yields a 36$\%$
error. Since the major role in the proposed combination is played
by the Mercury's node, it could happen that the new, more accurate
ephemerides available in future thanks to the BepiColombo mission
will offer an opportunity to improve the present unfavorable
situation.

\keywords{Relativity --
                Gravitation--
                Celestial Mechanics--
                Sun: fundamental parameters --
                Planets and satellites: general--
                Methods: miscellaneous
                 }
}

\titlerunning {The solar gravitomagnetic field and its possible measurement}

\maketitle

\section{Introduction}
\subsection{The Lense-Thirring effect}
According to the linearized weak-field and slow-motion
approximation of the General Theory of Relativity (GTR), valid
throughout the Solar System, the secular gravitomagnetic
Lense-Thirring precessions on the longitude of the ascending node
$\Omega$ and the argument of pericentre $\omega$ of the orbit of a
test particle freely orbiting around a central mass $M$ with
proper angular momentum $J$ are (Lense \& Thirring
1918)\eqi\dot\Omega_{\rm LT}=\frac{2GJ}{c^2
a^3(1-e^2)^{\rp{3}{2}}},\ \dot\omega_{\rm LT}=-\frac{6GJ\cos
i}{c^2 a^3(1-e^2)^{\rp{3}{2}}}.\label{leti}\eqf In \rfr{leti} $G$
is the Newtonian constant of gravitation, $c$ is the speed of
light in vacuum, $a,e$ and $i$ are the semimajor axis, the
eccentricity and the inclination, respectively, of the orbit to
the reference $\{x,y\}$ plane which coincides with the equatorial
plane of the central mass. Its spin $J$ is assumed to be directed
along the $z$ axis.
\subsection{Attempts to measure the Lense-Thirring effect in the Solar System}
Up to now, there is not yet any direct observational check of this
prediction of GTR which can be considered reliable and
undisputable.

Indeed, the only performed attempts to detect the Lense-Thirring
precessions of \rfr{leti} in the Solar System arena are due to I.
Ciufolini and coworkers (Ciufolini et al. 1998). They analyzed the
orbital data of the existing laser-ranged geodetic LAGEOS and
LAGEOS II satellites in the gravitational field of the Earth over
an observational time span of a few years. The claimed total
accuracy would be of the order of 20-30$\%$, but, according to
other scientists, such estimates would be largely optimistic (Ries
et al. 2003).

In April 2004 the extraordinarily sophisticated GP-B mission
(Everitt et al. 2001) has been launched. Its goal is to measure
another gravitomagnetic effect in the terrestrial gravitational
field, i.e. the precession of the spins of four superconductor
gyroscopes (Schiff 1960) carried onboard. The claimed accuracy
would be of the order of 1$\%$ or better. The experiment should
last one year.

Almost twenty years ago it was proposed to launch a third
LAGEOS-like satellite$-$the LAGEOS III/LARES$-$ and to analyze the
time series of the sum of the residuals of the nodes of LAGEOS and
LARES (Ciufolini 1986) or some other combinations of residuals of
the nodes and the perigees of LARES and both the existing LAGEOS
satellites (Iorio et al. 2002). The obtainable accuracy would
probably be of the order of 1$\%$. Mainly funding problems have
prevented, up to now, from implementing such relatively easy and
cheap mission. Recently, the possibility of measuring the
Lense-Thirring precessions of \rfr{leti} by means of the
Relativity-dedicated OPTIS spacecraft, which could be launched in
the same orbital configuration of LARES, has been considered
(Iorio et al., 2004).

The recently proposed LATOR (Turyshev et al. 2004) and ASTROD (Ni
et al. 2004) missions would be sensitive to the gravitomagnetic
part of the bending of light rays and time delay in the
gravitational field of the Sun.

Finally, it must be noted that, according to K. Nordvedt Jr.
(Nordvedt 2003), the multidecadal analysis of the Moon's motion
with the Lunar Laser Ranging (LLR) technique strongly supports the
existence of the gravitomagnetic force\footnote{According to
Nordvedt (2003), the Earth--Moon range is affected by
long--periodic harmonic perturbations of gravitomagnetic origin
whose amplitudes are of the order of 5 m and the periods are
monthly and semi-monthly. The amplitudes of the lunar motion at
both these periods are determined to better than half a centimeter
precision in the total orbital fit to the LLR data. } as predicted
by GTR, although in an indirect way. ``It often has been claimed
that the presence of gravitomagnetism within the total
gravitational interaction has not been experimentally confirmed
and measured. Indeed, different experiments have been under
development to explicitly observe the effects of this historically
interesting prediction of general relativity. But this
gravitomagnetic acceleration already plays a large role in
producing the final shape of the lunar orbit, albeit in
conjunction with the rest of the total equation of motion; the
precision fit of the LLR data indicates that gravitomagnetism's
presence and specific strength in the equation of motion can
hardly be in doubt. [...] It would be impossible to understand
this fit of the LLR data without the participation of the
gravitomagnetic interaction in the underlying model, and with
strength very close to that provided by general
relativity\footnote{Here the Eddington-Robertson-Schiff PPN
$\gamma$ parameter (Will 1993) is quoted.}, $\gamma=1$".

In regard of the possibility of measuring explicitly the
Lense-Thirring precessions of \rfr{leti} from the analysis of the
orbital motion of proof masses in the gravitational field of a
real rotating astronomical mass like the Earth, it must be pointed
out that the main problems come from the aliasing effects induced
by a host of classical orbital perturbations, of gravitational and
non-gravitational origin, which unavoidably affect the motion of
the probes along with GTR. In particular, the even zonal harmonics
$J_{\ell}$ of the multipolar expansion of the gravitational
potential of the central mass induce secular classical precessions
which, in many cases, are larger than the gravitomagnetic ones of
interest. Moreover, also the non-gravitational perturbations, to
which the perigees of the LAGEOS--like satellites are particularly
sensitive,  are another important source of bias. As we will see,
the approach proposed by Ciufolini (1996) and Iorio (Iorio 2002;
Iorio \& Morea 2004) in the performed or proposed tests with
LAGEOS and LAGEOS II consists of suitably designing linear
combinations $\sum c_i\delta\dot\Omega^i_{\rm obs}+\sum
k_j\delta\dot\omega^j_{\rm obs}$ of orbital residuals which are
able to reduce the impact both of the even zonal harmonics of the
gravitational field of the central mass and of the
non--gravitational perturbations. In general, the coefficients
$c_i$ and $k_j$ which weigh the various orbital elements in the
combinations are a compromise between these two distinct needs.
\subsection{Aim of the paper}
In this paper we wish to investigate the possibility of extending
the Ciufolini-Iorio approach in order to try to measure the
Lense-Thirring effect of \rfr{leti} in the gravitational field of
the Sun from interplanetary ranging measurements to some of the
inner planets of the Solar System. The relevant parameters are
listed in Table \ref{astropar}.

At this point it is important to clarify what is the current
approach in testing post-Newtonian gravity from planetary data
analysis followed by, e.g., the Jet Propulsion Laboratory (JPL).
In the interplay between the real data and the equations of
motions, which include also the post-Newtonian accelerations
expressed in terms of the various PPN parameters (Will 1993), a
set of astrodynamical parameters, among which there are also
$\gamma$ and $\beta$, are simultaneously and straightforwardly
fitted and adjusted and a correlation matrix is also released.
This means that the post-Newtonian equations of motion are
globally tested as a whole in terms of, among other parameters,
$\gamma$ and $\beta$; no attention is paid to this or that
particular feature of the post-Newtonian accelerations. This is
similar to the LLR approach outlined before. On the contrary, our
aim is just to try to single out one particular piece of the
post--Newtonian equations of motion, i.e. the gravitomagnetic
acceleration.

%
\begin{table}
\caption[]{Relevant astronomical and astrophysical parameters used
in the text. The value for the Sun's angular momentum $J$ has been
obtained from (Pijpers 2003). The planetary data can be retrieved
at \texttt{http://nssdc.gsfc.nasa.gov/planetary/factsheet/}. }
    \label{astropar}
\centering
\begin{tabular}{cccc}
            \hline\hline
Symbol & Description & Value & Units\\
            \hline
$G$ & Newtonian gravitational constant & $6.67259\cdot 10^{-11}$ & m$^3$ kg$^{-1}$ s$^{-2}$\\
$c$ & speed of light in vacuum & $2.99792458\cdot 10^8$ & m s$^{-1}$\\
$GM$ & Sun's GM & $1.32712440018\cdot 10^{20}$ & m$^3$ s$^{-2}$\\
$R$ & Sun's equatorial radius & $6.9599\cdot 10^8$ & m\\
$J$ & Sun's proper angular momentum & $1.9\cdot 10^{41}$ & kg m$^2$ s$^{-1}$\\
A.U. & astronomical unit & $1.49597870691\cdot 10^{11}$ & m\\
$a_{\rm Mer}$ & Mercury's semimajor axis & 0.38709893 & A.U.\\
$a_{\rm Ven}$ & Venus's semimajor axis & 0.72333199 & A.U.\\
$a_{\rm Ear}$ & Earth's semimajor axis & 1.00000011 & A.U.\\
$a_{\rm Mar}$ & Mars's semimajor axis & 1.52366231 & A.U.\\
$e_{\rm Mer}$ & Mercury's eccentricity & 0.20563069 & -\\
$e_{\rm Ven}$ & Venus's eccentricity & 0.00677323 & -\\
$e_{\rm Ear}$ & Earth's eccentricity & 0.01671022 & -\\
$e_{\rm Mar}$ & Mars's eccentricity & 0.09341233 & -\\
$i_{\rm Mer}$ & Mercury's inclination to the ecliptic & 7.00487 & deg\\
$i_{\rm Ven}$ & Venus's inclination to the ecliptic & 3.39471 & deg\\
$i_{\rm Ear}$ & Earth's inclination to the ecliptic & 0.00005 & deg\\
$i_{\rm Mar}$ & Mars's inclination to the ecliptic & 1.85061 & deg\\
            \hline
         \end{tabular}
\end{table}
%
\begin{table}
\caption{Gravitomagnetic and classical nodal precession
coefficients, in \asec. The coefficients $\dot\Omega_{.\ell}$ are
$\partial\dot\Omega^{J_{\ell}}_{\rm class}/\partial J_{\ell}$ and
refer to the classical precessions induced by the oblateness of
the central mass. The numerical values of Table \ref{astropar}
have been used in \rfr{leti} and \rfr{precclass} (see below).
$\dot\Omega_{\rm class}$ are the nominal centennial rates released
at \texttt{http://ssd.jpl.nasa.gov/elem$\_$planets.html}. They
mainly include the $N-$body secular precessions. } \label{prec}
\centering
\begin{tabular}{ccccc}
            \hline\hline
Precessions & Mercury & Venus & Earth & Mars\\
            \hline
$\dot\Omega_{\rm LT}$ & $1.008\cdot 10^{-3}$ & $1.44\cdot 10^{-4}$ & $5.4\cdot 10^{-5}$ & $1.5\cdot 10^{-5}$\\
$\dot\Omega_{.2}$ & $-1.26878626476\cdot 10^5$ & $-1.3068273031\cdot 10^{4}$ & $-4.210107706\cdot 10^3$ & $-9.80609460\cdot 10^2$ \\
$\dot\Omega_{.4}$ & $5.2774935\cdot 10^1$ & $1.349709$ & $2.28040\cdot 10^{-1}$ & $2.3554\cdot 10^{-2}$\\
$\dot\Omega_{\rm class}$ & $-4.4630\cdot 10^2$ & $-9.9689\cdot
10^2$ & $-1.822825\cdot 10^4$ & $-1.02019\cdot 10^3$\\ \hline
\end{tabular}
\end{table}
%
\section{The gravitomagnetic field of the Sun}
In the  case of the Sun and the planets, the Lense-Thirring effect
is quite small: indeed, for, e.g., the node it is $\leq$ $10^{-3}$
arcseconds per century (\asec), as it can be inferred from Table
\ref{astropar} and Table \ref{prec}. The point is that the angular
momentum of the Sun is relatively small and the Lense-Thirring
precessions fall off with the inverse of the third power of the
planet's semimajor axis. It is important to note that, if we want
to consider the detection of \rfr{leti} as a genuine test of GTR,
it is necessary that the Sun's angular momentum $J$ is known with
a high accuracy from measurements which are independent from GTR
itself. This is just the case: indeed, the helioseismic data from
the Global Oscillations Network Group (GONG) and also from the
Solar and Heliospheric Observatory (SoHO) satellite yield
measurements of $J$ which are accurate to a few percent (Pijpers
2003).
\subsection{Sensitivity analysis}\label{sensit}
According to the results of Table \ref{sensi} (E.M. Standish,
private communication, 2004), it should be possible to extract the
gravitomagnetic signature from a multi-year analysis of the
planetary nodal evolution. Let us explain how the results of Table
\ref{sensi} have been obtained. E.M. Standish averaged, among
other things, the nodal evolution of some planets over two
centuries by using the DE405 ephemerides (Standish 1998) with and
without the post-Newtonian accelerations. Standish included in the
force models also the solar oblateness with $J_{2}=2\cdot
10^{-7}$, so that the so obtained numerical residuals accounted
for the post-Newtonian effects only; the uncertainty in the
determined shift for, e.g., Mercury, was $
1.82\cdot 10^{-4}$ $''$ cy$^{-1}$. Note that the quoted
uncertainty of Table \ref{sensi} do not come from direct
observational errors. They depend on the fact that in the force
models used in the numerical propagation many astrodynamical
parameters occur (masses of planets, asteroids, etc.); their
numerical values come from multiparameter fits of real data and,
consequently, are affected by observational errors. Such numerical
tests say nothing about if GTR is correct or not; they just give
an idea of what would be the obtainable accuracy set up by our
knowledge of the Solar System arena if the Einstein theory of
gravitation would be true.
%
\begin{table}
\caption{Sensitivity in the numerical propagation of the planetary
nodal rates averaged over 200 years, in \asec. (E.M. Standish,
private communication, 2004)} \label{sensi} \centering
\begin{tabular}{cccc}
\hline\hline
Mercury & Venus & Earth & Mars \\
\hline
$1.82\cdot 10^{-4}$ & $6\cdot 10^{-6}$ & $2\cdot 10^{-6}$ & $1\cdot 10^{-6}$\\
\hline
\end{tabular}
\end{table}
It must be noted that our knowledge of the orbital motion of
Mercury will improve thanks to the future hermean missions
Messenger (see on the WEB http://messenger.jhuapl.edu/ and
http://discovery.nasa.gov/messenger.html), which has been launched
in the summer 2004 and whose encounter with Mercury is scheduled
for 2011, and, especially\footnote{While the spacecraft trajectory
will be determined from the range-rate data, the planet's orbit
will be retrieved from the range data (Milani et al. 2002). In
particular, the determination of the planetary centre of mass is
important to this goal which can be better reached by a not too
elliptical spacecraft's orbit. The relatively moderate ellipticity
of the planned 400$\times$ 1500 km polar orbit of BepiColombo,
contrary to the much more elliptical path of Messenger, is, then,
well adequate.}, BepiColombo (see on the WEB
http://sci.esa.int/science-e/www/area/index.cfm?fareaid=30), which
is scheduled to fly in 2010-2012. A complete error analysis for
the range and range-rate measurements can be found in (Iess \&
Boscagli 2001). According to them, a two orders of magnitude
improvement in the Earth-Mercury range should be possible.
According to a more conservative evaluation by E.M Standish
(Standish, private communication 2004), improvements in the
Mercury's orbital parameters might amount to one order of
magnitude;
Indeed, the current accuracy in radar ranging is hundreds of
meters; The new data could be reach the tens of meters level.
These figures allow to get another evaluation of the present and
future accuracy in measuring the Lense-Thirring precession on the
hermean node. The shift in the position of Mercury's node due to
the solar gravitomagnetic field over a time span $T$ is  $\Delta
r=a\dot\Omega_{\rm LT}T$; from the data of Table \ref{astropar} it
can be retrieved that, over one century, it amounts to 282 m. By
assuming $\sigma_{r_{\rm Merc}}\sim 100$ m we have a 35$\%$ error
which would reduce to 3$\%$ for $\sigma_{r_{\rm Merc}}\sim 10$ m.

On the other hand, there would be severe limitations to the
possibility of detecting the Lense-Thirring effect by analyzing
the secular evolution of only one orbital element of a given
planet due to certain aliasing systematic
errors\footnote{Fortunately, contrary to the Earth-LAGEOS-LAGEOS
II system, in this case the non-gravitational perturbations, which
are proportional to the area-to-mass ratio $S/M$ of the orbiting
probe, can be safely neglected. Indeed, $S/M\propto 1/r$ where $r$
is the radius of the probe assumed spherical.}. Indeed, as in the
case of the Earth-LAGEOS-LAGEOS II system, we should cope with the
multipolar expansion of the central mass, i.e. the Sun in this
case. Indeed, it turns out that the aliasing secular precessions
induced by its quadrupole mass moment $J_2$ on the planetary nodes
and the perihelia would be almost one order of magnitude larger
than the Lense-Thirring precessions if we assume $J_2=2\cdot
10^{-7}\pm 4\cdot 10^{-8}$ (Pireaux \& Rozelot 2003). There are
still many uncertainties about the Sun's oblateness, both from a
theoretically modelling point of view and from an observational
point of view (Rozelot et al. 2004). Moreover, the perihelia are
also affected by another relevant post-Newtonian secular effect,
i.e. the gravitoelectric Einstein pericentre advance (Einstein
1915)\eqi \dot\omega_{\rm GE }=\rp{3nGM}{c^2 a (1-e^2)}.\eqf In it
$n=\sqrt{GM/a^3}$ is the Keplerian mean motion of the orbiting
particle. This effect has been measured for the first time in the
Solar System with the interplanetary ranging technique at a
$10^{-3}$ level of relative accuracy (Shapiro et al. 1972; 1976;
Shapiro 1990). The Einstein precession is almost four orders of
magnitude larger than the Lense-Thirring effect, so that its
mismodelled part would still be one order of magnitude larger than
the Lense-Thirring effect of interest. Finally, also the classical
$N-$body secular precessions are to be considered because it turns
out that they are quite large (see Table \ref{prec}).
\section{The linear combination approach}
A possible way out could be to extend the Ciufolini-Iorio linear
combinations approach to the Sun-planets scenario in order to
built up some combinations with the nodes of the inner planets
which cancel out the impact of the Sun's oblateness and of the
$N-$ body precessions.
\subsection{A $J_2-(N-$body$)$ free combination}
Let us write down the expressions of the residuals of the nodes of
Mercury, Venus and Mars explicitly in terms of the mismodelled
secular precessions induced by the quadrupolar  mass moment of the
Sun, the secular $N-$body precessions and of the Lense-Thirring
secular precessions, assumed as a totally unmodelled feature. It
is accounted for by a scaling parameter $\mu_{\rm LT}$ which is
zero in Newtonian mechanics and 1 in GTR\footnote{It can be shown
that it can be expressed in terms of the PPN parameter $\gamma$.}
\begin{equation}\left\{\begin{array}{lll}
\delta\dot\Omega^{\rm Mercury}_{\rm obs}=\dot\Omega^{\rm
Mercury}_{.2} \delta J_{2}+\dot\Omega_{\rm class}^{\rm Mercury}
+\dot\Omega^{\rm Mercury}_{\rm LT}\mu_{\rm LT}+\Delta^{\rm Mercury},\\\\
\delta\dot\Omega^{\rm Venus}_{\rm obs}=\dot\Omega^{\rm Venus}_{.2}
\delta J_{2}+\dot\Omega^{\rm Venus}_{\rm class}
+\dot\Omega^{\rm Venus}_{\rm LT}\mu_{\rm LT}+\Delta^{\rm Venus},\\\\
\delta\dot\Omega^{\rm Mars}_{\rm obs}=\dot\Omega^{\rm Mars}_{.2}
\delta J_{2}+\dot\Omega_{\rm class}^{\rm Mars}+\dot\Omega^{\rm
Mars}_{\rm LT}\mu_{\rm LT}+\Delta^{\rm Mars}.\label{syst}
\end{array}\right.\end{equation}
The coefficients $\dot\Omega_{.{\ell}}$ are defined as
\eqi\dot\Omega_{.{\ell}}=\frac{\partial \dot\Omega^{J_{\ell}}_{\rm
class}}{\partial J_{{\ell}}},\nonumber\eqf where
$\dot\Omega^{J_{\ell}}_{\rm class }$ represent the classical
secular precessions induced by the oblateness of the central mass.
The coefficients $\dot\Omega_{.{\ell}}$ have been explicitly
worked out from $\ell=2$ to $\ell=20$ (Iorio 2003); it turns out
that they are functions of the semimajor axis $a$, the inclination
$i$ and the eccentricity $e$ of the considered planet:
$\dot\Omega_{.{\ell}}=\dot\Omega_{.{\ell}}(a,\ e,\ i;\ GM)$. For
the first two even zonal harmonics we have
\begin{equation}\left\{\begin{array}{lll}\label{precclass}
\dot\Omega_{.2}=-\frac{3}{2}n\left(\frac{R_{\odot}}{a}\right)^2\frac{\cos
i }{(1-e^2)^2},\\\\
\dot\Omega_{.4}=\dot\Omega_{.2}\left[\frac{5}{8}\left(\frac{R}{a}\right)^2\frac{\left(1+\frac{3}{2}e^2\right)}{(1-e^2)^2}
\left(7\sin^2 i-4\right)\right];
\end{array}\right.\end{equation}
 $R$ is the Sun's equatorial
radius. The quantities $\Delta$ in \rfr{syst} refer to the other
unmodelled or mismodelled effect which affect the temporal
evolution of the nodes of the considered planets. In the present
case they would mainly be represented by the precessions induced
by the octupolar mass moment of the Sun (Rozelot et al., 2004).
From the results of Table \ref{prec} and from the evaluations of
(Rozelot et al. 2004) according to which the possible magnitude of
$J_4$ would span the range $10^{-7}-10^{-9} $, the secular
precession induced by the octupolar mass moment of the Sun is
negligible with respect the Lense-Thirring rates. These facts
lead us to design a three-node combination which cancels out just
the effects of $J_2$ and of the classical $N-$body precessions
while is affected by the residual effect of $J_4$.

Indeed, if we solve \rfr{syst} for the Lense-Thirring parameter
$\mu_{\rm LT }$ it is possible to obtain
\begin{equation}\left\{\begin{array}{lll}\label{LTFORMULA}\delta\dot\Omega_{\rm obs}^{\rm
Mercury}  +  k_1\delta\dot\Omega_{\rm obs}^{\rm
Venus}+k_2\delta\dot\Omega_{\rm obs}^{\rm Mars}\sim X_{\rm
LT}\mu_{\rm
LT},\\\\
k_1=\frac{\dot\Omega^{\rm Mars}_{.2}\dot\Omega^{\rm Mercury}_{\rm
class } -\dot\Omega^{\rm Mercury}_{.2}\dot\Omega^{\rm Mars}_{\rm
class }}{\dot\Omega^{\rm Venus}_{.2}\dot\Omega^{\rm Mars}_{\rm
class }-\dot\Omega^{\rm Mars}_{.2}\dot\Omega^{\rm Venus}_{\rm
class}}=-1.0441702\cdot 10^1
,\\\\
 k_2=\frac{\dot\Omega^{\rm Mercury}_{.2}\dot\Omega^{\rm Venus}_{\rm class}
-\dot\Omega^{\rm Venus}_{.2}\dot\Omega^{\rm Mercury}_{\rm class
}}{\dot\Omega^{\rm Venus}_{.2}\dot\Omega^{\rm Mars}_{\rm class
}-\dot\Omega^{\rm Mars}_{.2}\dot\Omega^{\rm Venus}_{\rm
class}}=9.765758,\\\\X_{\rm LT}=\dot\Omega_{\rm LT }^{\rm
Mercury}+k_1\dot\Omega_{\rm LT}^{\rm Venus}+k_2\dot\Omega^{\rm
Mars }_{\rm LT }=-3.51\cdot 10^{-4}\ '' \ {\rm cy}^{-1},
\label{ltformula2}\end{array}\right.\end{equation} where the
numerical values of the coefficients $k_1$ and $k_2$ and the slope
$X_{\rm LT}$ of the gravitomagnetic trend come from the values of
Table \ref{prec}. The meaning of \rfr{LTFORMULA} is the following.
Let us construct the time series of the residuals of the nodes of
Mercury, Venus and Mars by using real observational data and the
full dynamical models in which GTR is purposely set equal to zero,
e.g. by using a very large value of $c$. We expect that, over a
multidecadal observational time span, the so combined residuals
will fully show the GTR signature and partly the mismodelled
$N-$body effect\footnote{Possible aliasing time-dependent $N-$body
residual effects with the periodicities of the outer planets,
mainly Jupiter, should average out over a sufficiently long
multidecadal time span. However, it would be possible to fit and
remove them from the time series.}in terms of a linear trend. The
measured slope, divided by $X_{\rm LT}$, yields $\mu_{\rm LT}$
which should be equal to one if GTR was correct and if the bias
from the residual $N-$body effect was sufficiently small. The
systematic error affecting \rfr{ltformula2} is totally negligible
because it would be due only to the higher degree multipole mass
moments of the Sun. The obtainable 1-sigma observational error,
according to the results of Table \ref{sensi}, would amount to
36$\%$. Note that this evaluation agrees with that presented in
Section \ref{sensit}.
\section{Conclusions}
In this paper we have explored the possibility of measuring the
post-Newtonian Lense-Thirring effect induced by the solar
gravitomagnetic field on the motion of some of the Solar System
planets. The magnitude of the gravitomagnetic precessions is very
small amounting to $10^{-3}$ \asec\ for Mercury. The main
systematic errors which would mask the relativistic effect of
interest would be the quite larger secular precessions induced by
the post-Newtonian gravitoelectric part of the Sun's gravitational
field, by the Sun's oblateness and by the $N-$body interactions.
By using a suitably designed linear combination of the orbital
residuals of the nodes of Mercury, Venus and Mars it would be
possible to cancel out the corrupting impact of the first solar
even zonal harmonic plus the $N-$body classical secular
precessions. Moreover, the proposed measurement would be aliased
neither by the post-Newtonian gravitoelectric field because it
affects only the perihelia and the mean anomalies. The obtainable
observable accuracy should be 36$\%$ (1-sigma) for the proposed
$J_2-(N-$body) free combination. It would be a somewhat modest
result for a reliable test of GTR. However, we note that should
new, more accurate ephemerides for Mercury be available as a
by-product from the Messenger and, especialy, BepiColombo
missions, the error's evaluation presented here could become more
favorable.
\begin{acknowledgements}
L. Iorio is grateful to L. Guerriero for his support while at
Bari. Special thanks to E. M. Standish (JPL) for his help and
useful discussions and clarifications, and to W.-T. Ni for the
updated reference about ASTROD.
\end{acknowledgements}


\end{document}